\documentclass{article}
\usepackage{fullpage}
\usepackage{url}

\begin{document}

\title{Thinking Taxonomically about Fake Accounts: Classification,\\False Dichotomies, and the Need for Nuance}

\author{
  Rebekah Overdorf\thanks{Order Alphabetical. Both authors contributed equally to this research.}\\
  EPFL\\
  \texttt{rebekah.overdorf@epfl.ch}
  \and
  Christopher Schwartz\\
  KU Leuven\\
  \texttt{christopher.schwartz@kuleuven.be}
}

\date{}


\maketitle

\begin{abstract}

It is often said that war creates a fog in which it becomes difficult to discern friend from foe on the battlefield. In the ongoing war on fake accounts, conscious development of taxonomies of the phenomenon has yet to occur, resulting in much confusion on the digital battlefield about what exactly a fake account is. This paper intends to address this problem, not by proposing a taxonomy of fake accounts, but by proposing a systematic way to \emph{think taxonomically} about the phenomenon. Specifically, we examine fake accounts through both a combined philosophical and computer science-based perspective. Through these lenses, we deconstruct narrow binary thinking about fake accounts, both in the form of general false dichotomies and specifically in relation to the Facebook's conceptual framework ``Coordinated Inauthentic Behavior'' (CIB). We then address the false dichotomies by constructing a more complex way of thinking taxonomically about fake accounts.

\end{abstract}

\section{Introduction}

The rapid rise of online social networks (OSNs) and online social media platforms has changed the way that audiences interact with journalism, news, and each other. This shift has led to a decentralization of news-reporting and information dissemination in general. Alongside the many advantages of this decentralization, however, come many disadvantages, key among them being the undermining of the trust model of traditional news, in which news institutions act as ``gatekeepers'' of information~\cite{gatekeeper_theory}. Essentially, the old ``one-to-many'' model of news has been replaced by a new many-to-many mode~\cite{many_to_many}. Meanwhile, platforms are reluctant to take on the duties of gatekeeping. As recently evidenced by the controversy surrounding Twitter's decision to begin flagging the tweets of United States President Donald Trump for violent, misleading, or false statements, when platforms become gatekeepers, explosive social-political consequences can follow~\cite{Trump_vs_Twitter}. 
All told, this has resulted in a situation in which misinformation and its more dangerous cousin, disinformation, can spread swiftly through an online audience.

Due to the proliferation of digital misinformation and disinformation it has become necessary to study a primary vector of their distribution within digital spaces: fake accounts. Research into fake accounts has primarily focused on detection and, increasingly, control. There have been two historic drivers of this: on the one hand, the influx of fake accounts into digital spaces, which has reached crisis proportions in recent years --- e.g., in May 2019, Facebook reported that it took down three \emph{billion} fake accounts, and five percent of its total monthly account activity was constituted by fake accounts~\cite{facebooktakedown} --- combined with the ease of Twitter data collection. 

In this paper, we take a critical look at the lack of conscious development of taxonomies around the phenomenon of fake accounts and propose a systematic way to \emph{think taxonomically} about the phenomenon. Specifically, we combine methods from computer science and philosophy to create a comprehensive theory of fake accounts, including definitions of what it means to be ``fake'' and a fake account, and key taxonomical aspects of the latter. Along the way, we deconstruct the narrow binary thinking surrounding fake accounts as specifically exhibited in Facebook's conceptual framework of ``Coordinated Inauthentic Behavior'' and as more generally manifested in a series of false dichotomies about the phenomenon. 

\paragraph{Taxonomical Thinking}

Taxonomies must strike an optimal but difficult balance between mixed, or even opposing, intuitions and methods of analysis about the phenomenon they are intended to typologize or categorize. The most effective taxonomies are those which resist two major temptations: a) binary frameworks and b) over-reliance on either quantitative or qualitative methods of analysis. Along these lines, taxonomical thinking is a meta-level activity that occurs prior to (or simultaneously with) the development of an actual taxonomy. The would-be taxonomist should consciously think through not only the necessary typologies or categories, but the meta-types or meta-categories themselves, viz., the quantifiable and qualitative \emph{aspects} --- the empirical elements and the intuitions --- that the types or categories are intended to capture.

Taxonomical thinking is, at root, a philosophical endeavor. Taxonomies are implicit ontologies, for every taxonomist, consciously or unconsciously, strives to answer the ontological question, \emph{What is X?} In the case of fake accounts, any taxonomizer must first think through \emph{two} ontological questions --- what is ``fakeness'' and what are fake accounts? --- \emph{before} attempting to answer the typological question --- which types of fake accounts are there?

\paragraph{Classification as Substitute Taxonomization}

In the brief history of research into fake accounts, there has been a proliferation of taxonomies under the guise of machine learning-based classifiers developed and deployed to detect, control, and study the phenomenon. Supervised machine learning classification is the process of partitioning data into classes based on a set of labeled ground truth data points, while taxonomies are mental devices that likewise partition objects of experience into taxa based on a set of categorized concrete examples. Both supervised machine learning and taxonomical development work under the premise that samples/objects can be categorized based on the similarities and differences of their features. While not always equatable at the level of operation, classification and taxonomization are equatable at the level of premises, starting-points (i.e. data), and goals. They are two sides of the same coin. It stands to reason, then, that classifiers often act, if implicitly, as taxonomists.

The use of classification as not even a proxy for taxonomization, but a substitute for it, is deeply problematic. Put simply, we tend to over-rely on quantitative methods of analysis of fake accounts in order to skirt trickier, slippery qualitative methods and questions. The result is that, as is common in quantitative analysis, researchers derive binary conclusions about fake accounts, even when multiple classes are present. These binary conclusions often lead to false dichotomies. We identify and deconstruct four such dichotomies in this work: i) coordination versus non-coordination, ii) program versus program, iii) deception versus forthrightness, and iv) inauthenticity versus authenticity.

\paragraph{CIB as a Case Study}
The need for conscious, critical taxonomical thinking that incorporates qualitative modes of analysis with quantitative ones is most evident in Facebook's CIB framework. Presumably, CIB is not just a theoretical construct, but is intended to shape Facebook's internal policies about fake accounts~\cite{CIB}. We build on the critique of CIB by Starbird et al.~\cite{starbird} and find that the problems of Facebook's framework go beyond coordination versus collaboration; rather, it assumes its own awareness of descriptive, security, and normative concerns. As a result, CIB makes it difficult for everyone involved in this struggle --- researchers, journalists, law enforcement and national security officers, audiences of everyday users, policy-makers, and platform managers --- to grapple with the vast range of permutations of fake accounts.

\paragraph{``Innocent'' Fake Accounts} Crucially, CIB also obscures the role played by intention. Simply put, as we demonstrate in this paper, there is no sound philosophical basis to believe that fake accounts are inherently and invariably malicious, even if there is some degree of inherent capacity for maliciousness due to the fact that fake accounts are inevitably a form of deception. In other words, users of fake accounts are not always and consistently seeking ``to falsify or mislead for purposes beneficial to the agent and harmful to the target''~\cite{Stockholm_2019}. They are not, then, \emph{ipso facto} adversaries. 

That every user of a fake account is not inherently intending to commit adversarial acts creates a problem that transcends classification and which can only be addressed at the level of taxonomization. Essentially, researchers need to be able to infer intention from observable data and engineer features for it so that classifiers can input for it. However, this is both infeasible given current technology, as there are no known reliable indicators for intention within platforms, and even were it possible to establish inferences, it may be morally dubious. Consequently, making decisions about interpretation purely on the basis of classifier outputs is scientifically incomplete, intellectually faulty, and morally questionable.

\paragraph{Road Map} This paper will proceed as follows. Section \ref{sec:related} will provide context for both the problem of using classification as an implicit form of taxonomization and for the inclusion of more qualitative modes of analysis (in particular, in the form of philosophy). Section \ref{sec:adversarialness} follows with a critique of considering \emph{inauthenticity} and \emph{coordination} as primary factors of fake accounts. This critique will take the form of a philosophical thought experiment. Section \ref{sec:falsedichotomies} then philosophically analyzes four false dichotomies tied to binary thinking about the problem of fake accounts. Section \ref{sec:taxonomy} presents a system of thinking taxonomically comprised of aspects, each corresponding to a false dichotomy. Section \ref{sec:inpractice} then explores example case studies that demonstrate the use and utility of our proposed aspects. Finally, section \ref{sec:conclusion} concludes with some gestures toward future work.

\section{Context and Rationale}
\label{sec:related}

In this section, we make the case for why taxonomical thinking is required to grapple with fake accounts. We will begin by comparing taxonomization and classification to elucidate how classifiers frequently double as implicit taxonomies. Following this, we will provide examples from the literature where classification has served in all but name as taxonomization. We will then justify the decision to incorporate philosophical analysis into the discussion of taxonomizing fake accounts.

\subsection{Classification as Implicit Taxonomization}

That classification is implicitly taxonomization may seem obvious due to the fact that they clearly have the same end goal: to categorize. However, there may also be resistance to this claim on the basis that, while a comparison is indeed valid, classification and taxonomization are not \emph{equatable} as actual activities. To see why they are equatable, we must take a moment to, as it were, go ``under the hood'' and show how classification and taxonomization's premises and starting points are in fact the same, even if the way that they operate is indeed not always identical.

Supervised classification in machine learning is the process of partitioning data into classes based on a set of labeled ground truth data points. Decisions about which class to sort each sample into are made based on the features of each sample. That is, a supervised machine learning algorithm aims to find a function $g:X \rightarrow Y$ where $X$ is the input space (features) and $Y$ the output space (labels), given $N$ training samples $\{(x_0,y_0),(x_1,y_1)...,(x_n,y_n)\}$.

Taxonomies are mental devices that likewise partition objects of experience into taxa based on a set of categorized concrete examples. Decisions about which taxon to sort each example into, however, depend on several factors. To put this in terms of the formalization above, a taxonomy is a mapping, $g:X \rightarrow Y$, where $X$ is the input space (features) and $Y=$ the output space (taxa) that is developed given $N$ concrete examples: $\{(x_0,y_0),(x_1,y_1)...,(x_n,y_n)\}$. That is, the taxonomy itself can be considered as $g$, the method for mapping an object of experience (based on its features $x_i$) to a taxon in $Y$. 

Both supervised machine learning and taxonomical development work under the premise that samples/objects can be categorized based on the similarities and differences of their features. Take for example an image recognition task, in which a classifier is trained on a set of features involving animals and asked to classify new images of animals. The classifier relies on the fact that, e.g., all penguins (in the training set) are black and white. Similarly, a taxonomy of animals can be developed to categorize animals according to visual observations, e.g., all male lions (that its author has observed) have luxurious manes.

Binary classification is perhaps the most obvious example in which classification and taxonomy can be equated. Binary classification concerns two classes, $Y=\{y,\neg{y}\}$. The classifier is trained on samples from each class $\{(x_0,y),...,(x_m,y),(x_{m+1},\neg{y}),...,(x_n,\neg{y})\}$, which allows the classifier to learn how to identify which class a new sample belongs to. Similarly, binary taxonomies strive to do two things simultaneously on behalf of the taxonomist: a) encompass as many instances and permutations as possible of that which they are categorizing ($y$), while also b) delineating as clear conceptual borders as possible to designate that which they are \emph{not} categorizing, or what should be categorized as not relevant ($\neg{y}$). These two efforts can be considered as any given taxonomy's positive and negative categorizing principles, respectively. A philosophical example of a binary taxonomy would be ``self'' versus ``other'' (the latter understood as ``\emph{not} self'').

Additionally, both classification and taxonomization can introduce hierarchy valuations either within and/or between their resulting sets. Using binaries again as the example, hierarchy valuations are essentially designating the $y$ as somehow of higher status than $\neg{y}$. Taxonomies perform hierarchy valuations frequently, designating things as ``bad'' or undesirable, as in taxonomies of justice which distinguish between ``just'' versus ``unjust'' outcomes, the former being intuitively preferred over the latter. Within classification, there is sometimes the belief that no hierarchy valuations \emph{per se} are occurring, such that in binary classification, the $y$ and $\neg{y}$ are simply descriptive and do not carry any normative judgments. Yet, even as simple a binary classification as ``dog'' versus ``cat'' can be imbued with a hierarchy valuation if the classifier is also trying to find dogs. Finding cats, as the ``not dog'', is subordinated to that goal.

Note that if a taxonomy does not have a clear negative principle ($\neg{y}$), then it can behave in other ways. For example, objects that maximally resemble its key categories may clustered together at one end of a spectrum, while those with minimal resemblance may be spread out in a long tail or clustered at the opposite end. In the classical example from Aristotle, the ancient philosopher does not categorize plants and fungi as animals, but he does identify many shared features, e.g., they are organic, they have metabolisms, and they can move (although not to the same degree as creatures with legs)~\cite{De_Anima}. Following Aristotle's observation, if placed on a spectrum, plants and fungi would be clustered nearer to animals than viruses, while viruses would themselves also be nearer to animals than stones. 

While it may be tempting to compare spectrum-style taxonomization to a method of classification such as k-Nearest Neighbors (kNN), it is really only a superficial resemblance. This leads us to repeat an important point: classification and taxonomization are equatable at the levels of premise, starting points, and goals, but not necessarily at the level of operation. An even stronger example of divergence than kNN are Neural Networks: with their vast weaves of input/outputs and weights, they do not resemble taxonomies at all. This is also why, however, it is important that classification not be used as a substitute for taxonomization. They are two sides of the same categorizing coin, but they are still two sides.

\subsection{Incorporating Philosophical Analysis}

Classification is not a reliable substitute for thorough and \emph{deliberate} taxonomical thinking for two reasons. First, it feeds into a very human temptation to suspend thinking and let someone or something else carry the burden of making decisions about interpretation. Second, the resulting interpretations themselves are over-reliant on quantitative methods of analysis and hence cannot truly grapple with the complex and subtle challenge posed by fake accounts. 

Machine learning does not and \emph{should not} ``make'' decisions about interpretation. Further below in this paper, we will explore four false dichotomies that arise when machine learning research outcomes about fake accounts are interpreted as binary $y$ and $\neg{y}$ conceptual frameworks. The most obvious example of this will be person versus program (``bot or not'').

Meanwhile, the over-reliance on quantitative methods of analysis simply cannot grasp the slippery dynamics of fake accounts. In a recent blog post, Twitter itself has made this observation: 

\begin{quote}
``What’s more important to focus on in 2020 is the holistic behavior of an account, not just whether it's automated or not. That's why calls for bot labeling don't capture the problem we're trying to solve and the errors we could make to real people that need our service to make their voice heard. It's not just a binary question of bot or not --- the gradients in between are what matter''~\cite{twitterblog}.
\end{quote}

Take for example an OSN audience witnessing a debate between two accounts that, unbeknownst to them, are secretly controlled by one person. The puppeteer's goal is to persuade the audience to believe in position $A$ and dissuade them from believing in position $B$. Crucially, position $A$ would be beneficial to the audience's well-being, while position $B$ would be detrimental (e.g. anti-vaccines). Now, imagine if one of the audience-member's users grows suspicious of the two interlocutors. This would-be sleuth deploys stylometric analysis on their posts and determines with a high probability that the two accounts are one person and then alerts their fellow audience-members to the deception. However, to the surprise and perhaps horror of the sleuth, the others decide that it is not important whether the two interlocutors really are two people; rather, what matters is the content of their debate.

Such a scenario poses a problem for a classifier that is also implicitly doubling as a taxonomist: it can identify the two accounts as ``sockpuppets'', but given what is known about both the intention of the puppeteer and especially the audience's reaction to learning about the two accounts' fakeness, there seems something \emph{intuitively} insufficient in this categorization. Of course, categories are made to be parsed and subdivided. However, what is important is whether the inevitable and necessary parses and subdivisions are based solely on quantitative taxa, such as accounts' scale, characteristics, and behaviors, or also more qualitative taxa, such as inferences about their hidden owners, their intentions and the purposes of the accounts themselves, audience reactions to the behaviors of the accounts, etc.

The enigmatic nature of qualitative taxa often discourages researchers from even attempting to discern them, which in turn fosters the tendency to use classifiers as implicit taxonomists and make decisions about interpretation on quantitative methods alone. To be sure, some researchers, intuiting the limits of such an approach, have attempted to address it~\cite{lit_rev_fake_accounts, attempt1, attempt2}. However, they have largely continued to do what philosophers would call ``mathematizing the problem'', viz., seeking quantitative manifestations of qualitative things. In other words, they engineer better and more subtle features derived from ever-more granular empirical indicators, while the labor of conscious and purposeful taxonomization remains to be done.

In fairness, to a great extent reliance on quantitative methods and empirical indicators for non-empirical realities is inescapable in computer science. Moreover, there will always be an interplay between the qualitative and quantitative dimensions of life, such that there \emph{should} be sustained efforts to discern subtler and more meaningful empirical indicators for non-empirical realities. Yet, for precisely this reason, an interdisciplinary approach that can incorporate qualitative methods of analysis is required. Technology, as a field, simply cannot grapple with fake accounts on its own.

Among qualitative methods, the philosophical tradition proves to be the most useful. Since Aristotle's \emph{Categories}, philosophers have been developing theories of taxa, taxonomies and, most of all, taxonomical thinking. More crucially, philosophy has been many things over the centuries, but first and foremost it has been a discipline focused on honing the very meta-level of observation and analysis that taxonomizing fake accounts requires. By way of illustration, our observation here that classification is implicit taxonomization is, at root, a philosophical observation.

\section{Problems of Coordinated Inauthentic Behavior}

\label{sec:adversarialness}

To demonstrate why a lack of conscious taxonomical thinking is problematic, here we pose a thought experiment that illustrates the limits of Facebook's CIB framework. We take our thought experiment from Orson Scott Card's science fiction novel \emph{Ender's Game}~\cite{Ender}. 

To be sure, Facebook is attempting to describe a legitimate intuition shaped by real and frightening instances of disinformation attacks from recent history~\cite{russiantrolls, brexittrolls, spanishtrolls}. In this respect, CIB is a potentially useful way to describe a \emph{specific} subset of fake accounts and their behaviors. The problem is that it is insufficient to describe fake accounts and their behaviors \emph{as such}.

In Card's novel, the sibling prodigies Valentine and Peter Wiggin use two sockpuppet bloggers called ``Demosthenes'' and ``Locke'' to bring about world peace through a fake philosophical conversation. Demosthenes has a militaristic character, while Locke is peaceful; the former's true purpose is to gradually make latter appear as the more persuasive speaker. Eventually, the Wiggins are revealed as the puppeteers. However, this revelation does not perturb their worldwide audience, who proceed to install Peter as ``hegemon'' of the planet~\cite{Ender}.

For the sake of argument, we tweak Card's scenario in three ways. First, Demosthenes becomes an ideological Right-winger, Locke an ideological Left-winger, and neither is necessarily militaristic nor pacifistic. Second, they remain rhetorically evenly matched and achieve a consensus in which their two perennially-opposed social-political persuasions are finally reconciled on equal terms. Third, the Wiggins do not reveal themselves; instead, the way in which the two sockpuppets achieve their consensus is so incredibly persuasive for their audience that everyday Internet users become convinced to spread it to their friends and families of their own free will. In doing so, they steadily increase the Wiggins' reach and impact until even governments succumb to the persuasiveness of the two sockpuppets. Ultimately, their universal audience of its own accord establishes a new global order that integrates Right- and Left-wing understandings of society.

Both Card's original scenario and our revised one involve coordinated inauthentic behavior. Yet, in both scenarios the audiences are unperturbed by the deception. In Card's original, the deception is revealed, while in our revision, the audience may infer it from the fact that ``Locke'' and ``Demosthenes'' are obviously pseudonyms, their namesakes long dead. In both scenarios, the audience psychologically minimizes the fact of this deception, instead assigning greater value to the content of the two sockpuppet's arguments. 

The conceptual framework of CIB, by making coordination and inauthenticity its centerpieces, makes it difficult to comprehend these scenarios. Crucially, CIB assumes that it knows what are the descriptive, security and normative concerns at stake here. The behavior between the Wiggins' sockpuppets is indeed coordinated, but is it \emph{malicious}, much less inauthentic? We, Card's readers, are never really given insight into Peter's intentions, but it does appear that Valentine sincerely believes that consensus is necessary for the general well-being of society but that it can only be achieved via deception.

Starbird et al.~\cite{starbird} help us see that the Wiggins' deception is also not coordinated. It initially lacks scale: their scheme hinges entirely on only \emph{two} fake accounts. As things unfold, their audience takes over, actively attending to Demosthenes and Locke's conversation and participating in it via comments, their own blogs, and telling others about it.

In his exploration of Card's scenario, Kenneth Wayne Sayles III criticizes the Wiggins' audience as ``naive and foolhardy''~\cite{Ender_Rhetoric}. However, again following Starbird et al.'s analysis, the contrary seems more likely: the audience are \emph{more or less} witting accomplices to the Wiggins. In the original, when the Wiggins are revealed and the audience not only does not care about the deception, but rewards the siblings by installing Peter as ruler of the world. In our revision, the audience never learns the true identities of Demosthenes and Locke, but again, the fact of the deception, which they can intuit, does not perturb them. 

All told, Facebook's CIB framework has great difficulties contending with fake account's impacts on and implications for audiences. Are the Wiggins' audiences \emph{coordinating}? In so doing, are they themselves being \emph{inauthentic}? At an intuitive level, these questions miss the mark.

Even more fundamentally, the point of our thought experiment is this: the only correct classifier for fake accounts should have the account user's intention as an input, for the Wiggins had benign intentions even though they behaved deceptively. Unfortunately, such an input is both infeasible given current technology, as intention is not observable within a platform, and morally dubious even if there were indicators for it. Consequently, making decisions about interpretation purely based on classifier outputs is incomplete and hence intellectually faulty and morally questionable.
\section{False Dichotomies}
\label{sec:falsedichotomies}

Quantitative methods and classification techniques tend to lead to rather binary thinking. Take for example a problem quite distinct from fake account detection: hate speech detection~\cite{hatespeech}. Often, hate speech is considered as a binary: something either is hate speech or it is not. Yet, in reality, hate speech exists on a spectrum and a decision is made, whether implicitly or explicitly, about the location of the decision boundary. Binary thinking can lead to a hardline reaction: it must either be blocked/removed or left alone. Unfortunately, hardline reactions do not leave conceptual space to think through more proportional and nuanced solutions.

Binary thinking typically expresses itself philosophically through false dichotomies. With respect to fake accounts, we identify four false dichotomies: i) coordination vs. non-coordination, ii) program vs. person, iii) deception vs. forthrightness, and iv) inauthenticity vs authenticity. 

\subsection*{False Dichotomy 1: Coordination versus Non-Coordination}

In CIB, fake accounts are such at least partially because they are coordinated. This implies that ``real'' accounts are such at least partially because they are \emph{not} coordinated. Unfortunately, coordination rests on an intuition that seems obvious, but on philosophical inspection is misleading. 

Philosophically, coordination is the expectation that fake accounts typically entail two dimensions: a campaign and a network. The mental image is of a top-down, centralized, and cunningly-planned campaign with seamless and systematic cooperation between its various components. These components include not only fake accounts, but also content and other elements. ``Fake news'' is the pertinent example here, which can even include entire websites purporting to be news agencies that do not actually exist~\cite{fakenewssites}.

Coordination is one of the two bedrock dichotomies of Facebook's framework of CIB. Starbird et al.~\cite{starbird} have criticized CIB precisely for its emphasis on coordination. The authors find the dichotomy to be an oversimplification, since often actors are not explicitly cooperating or are even unaware of their role in a disinformation campaign. Misinformation patterns and disinformation operations can exhibit far greater and \emph{sloppier} human agency than the mental image that coordination suggests. Misinformation and disinformation also involve, and often thrive, on human error and the haphazard social dynamics of audiences.

Starbird et al. instead propose collaboration. Collaboration invites a broader understanding of how fake accounts interact not only with each other, but also with their real counterparts. Audiences are not passive recipients of the actions of fake accounts; rather, they can sincerely and well-intentionally interact with fake accounts for their own purposes. Crucially, audiences can suspect that some accounts may be real, or at least have an agenda, but disregard or downplay this. By contrast, the interactions among fake accounts themselves do not always entail purposeful and planned coordination. In general, the interactions between fake and real accounts, as well as between fake accounts, entails a ``collaboration consist[ing] of convergent behaviors that reflect a kind of intentional shaping or cultivation of an online community''. Moreover, ``that shaping is not uni-directional, but has elements of improvisation as the operators work to both reflect and guide the activities of their host community.''~\cite{starbird}

\subsection*{False Dichotomy 2: Program versus Person}

Program versus person concerns who or what a fake account's user may ``really'' be, and it is so intuitive as to seem common-sensical. There is likewise a presupposition at work here about the definition of fake versus real: fake accounts are controlled by programs, real accounts by people. 

Now, consider the Ship of Theseus, an ancient logical paradox from the philosophical tradition. Thousands of years ago, a wooden ship disembarked from Athens bearing precious gold in its cargo, with Alexandria its destination and the great hero Theseus its captain. The journey proved arduous: the ship was beset by pirates and storms, necessitating multiple stops among the Aegean islands. By the time it docked at Alexandria, the ship had undergone a dramatic transformation: all of its wooden planks, masts and sails had been replaced; its original crew all lost to the perils of the voyage, replaced by a mix of new recruits and captured pirates; the gold had been traded for gems; even Theseus himself was gone, a captain of another ship, the new captain hired from a fishing boat to complete the mission. Was the chimera of a vessel that arrived in Alexendria truly the ship that disembarked from Athens?~\cite{theseus}

Fake accounts pose a similar challenge as the ship of Theseus~\cite{theseusbots1,theseusbots2}. Imagine a sockpuppet created by a user to help their real account win a debate with another user gradually becomes the person's true account, not in nominal identity but in actual everyday use; slowly, the real account is abandoned. Later, it is hacked and assimilated into a ``bot army'' controlled by an algorithm. This bot army is then rented out to various actors, switching from broadcasting advertisement content to propagating disinformation to following an ``influencer''. At some point, the account's profile photograph and name become copied and recycled for entirely fabricated accounts, one taking the name, the other taking the photograph. 

In a sense, the history of fake accounts is moving in the direction not of a Ship of Theseus, but a Cyborg of Theseus. The sockpuppet and the bot are two social-digital identities whose purpose is to deceive audiences of other users. The sockpuppet is traditionally controlled by a person, while the bot is by definition controlled a program. However, they both pretend to be a real person \emph{other} than the person or program controlling them. For this reason, the sockpuppet and the bot are sometimes treated as the ``archetypes'' of fake accounts.

Sockpuppets are used to either praise, defend, or otherwise support a person or organization, often for the purpose of manipulating an audience's opinion. They pose as an independent third-party entity wholly unaffiliated with its ulterior user~\cite{classical_sockpuppets, gamers, cybercriminals}. However, depending on what the adversary is seeking to accomplish, sockpuppets can be inefficient: they depend on real people manually creating and managing them, and ideally, they must constantly be tended to produce content in order to appear authentic. Their scalability is ultimately a function of the number, efficiency and, most elusively, creativity of real people. In contrast, bots are easily scalable since there is less perceived need to make them imitate real people. For an adversary seeking a cost-effective way to manipulate audiences, a believable name and profile picture is sufficiently deceptive, as the sheer amount of bots following each other and sharing each other's content can camouflage their superficiality to unobservant users.

Despite the prevalence of bots on the Internet today, sockpuppets remain attractive to adversaries for two fundamental reasons. To begin with, bots' scalability makes them more susceptible to machine learning-based methods of detection. Since a bot is automated, it has tell-tale features that a sockpuppet does not, e.g., appearing \emph{ex nihilo} and/or \emph{en masse} within a platform, broadcasting the same piece of content \emph{verbatim} and simultaneously, and brute-forcing the same hashtag across newsfeeds~\cite{Vosoughi}. To be sure, bots can be made to mimic people more effectively, such as by simulating circadian rhythms and not acting simultaneously~\cite{stella2018bots}. Sockpuppets are also somewhat susceptible to machine learning-based methods of detection, in particular by training models to discern the behavioral and stylistic idiosyncrasies specific to the ulterior users~\cite{CyberSafety2020, Stylometry}. However, the features of this or that sockpuppet do not potentially expose all sockpuppets \emph{as such} the way that the features of one bot army can potentially expose all bot armies everywhere. So long as a sockpuppet's ulterior user is trained to avoid both their own idiosyncrasies and the tell-tale signs of bots, as well as to use proxy servers or virtual private networks, they can adeptly elude detection.

Sockpuppets are also better fit than bots for the goal of persuasion~\cite{CyberSafety2020}. From the perspective of an adversary, bots are a hammer, sockpuppets a scalpel. They manipulate audiences in very different ways: bots flood platforms with content, while sockpuppets build relationships of trust~\cite{CyberSafety2020}. Consequently, bots are more effective at \emph{censoring}, while sockpuppets are more effective at \emph{persuasion}. This divergence is rooted in their different fundamental abilities: sockpuppets can generally engage with other users more or less one-on-one, whereas bots can generally only engage with other users as a mass.

However, sockpuppets and bots have also been intersecting for quite some time as adversaries learn how to increase the former's efficiency, whether through employment or automation. For example, Wikipedia has struggled with the phenomenon automated sockpuppet editors for the better part of a decade~\cite{wiki}, while national security agencies have been implicated in using ``sockpuppet armies'' in various intelligence, counterintelligence and paramilitary operations~\cite{duterte,israel,GCHQ}. The point is: insofar that the sockpuppet and the bot are \emph{interpreted} as representing a clean dichotomy between fake accounts that are controlled by people or programs, then this interpretation is woefully outdated and false.

\subsection*{False Dichotomy 3: Deception versus Forthrightness}

This dichotomy wields together two distinct but intertwined intuitions, one about the agenda or purpose of fake accounts, the other about their behaviors or techniques. Both intuitions can be encapsulated in the broad terms of ``deception'' versus ``forthrightness''. Essentially, there is an expectation that fake accounts, by virtue of the fact that their \emph{nature} is deceptive --- they are, after all, purporting to be someone that they are actually not --- also have the \emph{goal} of deceiving. Accordingly, they will also use nefarious methods to go about their deception. As with the previous dichotomies, there is an ambiguous presupposition about the definition of real accounts, viz., they are forthright and do not deceive, neither in terms of their nature --- who they purport to represent is who they actually are --- nor in terms of their goals --- they seek to innocently socialize, entertain, communicate, etc. --- nor in terms of their behaviors --- they innocently socialize, entertain, communicate, etc.

Of the many philosophical problems this dichotomy falls afoul of (to say nothing of messy everyday human realities, in which deception and forthrightness constantly co-occur), perhaps the most crucial is David Hume's distinction between ``is'' and ``ought''~\cite{Hume}. There are different ways to formulate Hume's distinction, but for our purposes here, his insight is that what an entity \emph{ought} to do does not necessarily reflect what an entity \emph{is}. Conceptually, this distinction enables understanding that there is not a necessary causal connection between an entity's nature and its goals. Along these lines, just because a fake account's nature is deceptive does not \emph{necessarily} mean its goals are likewise deceptive. Moreover, a deceptive nature may not be inherently problematic from a moral perspective.

There appear to be many instances in which audiences seem to accept the deception of a fake account. For example, some ``chatbots'' are designed to simulate a person and are increasingly found on customer service websites~\cite{chatbots1}. Their identity is fake, yet the deception --- which can be easily discerned by either their awkward use of language, their circular and menu-like way of interacting, and the fact that they eventually bring into the chat a human customer service representative --- is experienced as annoying but acceptable, even useful, by many audiences~\cite{chatbots2}. Similarly, satirical accounts, which pose as popular figures\footnote{\url{https://twitter.com/DarthVader}}, concepts\footnote{\url{https://twitter.com/fakescience}} or even God\footnote{\url{https://twitter.com/TheTweetOfGod}}, do not arouse controversy --- to the contrary, their deception seems intrinsic to the way in which audiences enjoy them.

\subsection*{False Dichotomy 4: Inauthenticity versus Authenticity}

All of the foregoing leads us to the fourth and possibly most important, but also most intricate, of the false dichotomies concerning fake accounts: inauthenticity versus authenticity. Alongside coordination, inauthenticity is the other bedrock dichotomy of Facebook's CIB dichotomy. Inauthenticity implies an understanding of fakeness as a form of deception that is inherently malicious. Yet, there exist non-malicious forms of fakeness such as users creating alternate accounts for entertainment~\cite{wani18}. Some forms of fakeness are actually necessary for safe and secure personal exploration and development within digital spaces, and along these lines, there is a strong argument to be made that non-malicious fakeness can ultimately contribute to the overall well-being of society~\cite{elastic_self}. There are deeper questions at stake than simply whether an account is fake, but whether the intention behind a fake account is malicious.

Often identified with the philosopher Søren Kierkegaard~\cite{Kierkegaard1}, inauthenticity occurs when something either has a disputable origin and/or it is not a reliable, accurate representation of that which it purports to represent~\cite{authenticity}. The suspicion of malice arises within the concept of inauthenticity from the fact of deception, not the intention of the deceiver.

To illustrate, Kierkegaard is often taken as the proponent of the popular cultural notion that ``one must be true to oneself'', in the sense that a person must obey their innermost moral promptings, and not doing so constitutes a moral hazard to themselves and others~\cite{authenticity}. It follows, then, that authenticity occurs when something's origins are indisputable and that it is in fact an accurate representation of that which it purports to represent. Thus, if a person does not believe in God but regularly attends church or mosque, they are being inauthentic, i.e., their action does not originate from genuine faith, they do not accurately represent their true belief, and they risk undermining the integrity of themselves and the religious community. However, if they refuse to attend church or mosque, they are being authentic.

Unfortunately, inauthenticity is one of those concepts that sounds clear in the abstract but becomes murky in real life. The concept must concede ground to situations like a person applying cosmetics to improve their natural appearance. Such a situation seems to ``bend the rules'' of authenticity without breaking them. Yet, inauthenticity cannot contend with a more devilsome paradox experienced quite poignantly by every person at some point in their life: when one has \emph{two} equal, sincere but antithetical moral promptings. For example, a soldier who vows to both protect life \emph{and} to kill that which threatens life.

Fake accounts pose a similar complication for inauthenticity. For one thing, inauthenticity implies lying, but it is also possible to be inauthentic by exaggerating. To quote the philosopher Albert Camus: ``Lying is not only saying what isn't true. It is also, in fact especially, saying more than is true, and, in the case of the human heart, saying more than one feels''~\cite{Camus}. For another thing, a paradox: it is possible for a person to be inauthentic in order to be authentic. As Graham Parkes puts it, ``the choice of [wearing] a particular mask signifies a part of the person and can bring out a side of the wearer's personality that would otherwise remain hidden''~\cite{parkes1987facing}.

What Parkes means can be grasped if we look at the closely-related concepts of pseudonymity and allonymity. The former is the practice of using a ``false name'', the latter of using a person's real name by another. As practices, pseudonymity and allonymity raise the question of whether identity fraud is \emph{inherently} immoral. Consider the following points. First, pseudonymity and allonymity are, essentially, acts of masquerade and imposture, respectively. Masquerade, whether literal or metaphorical, is the nature of ``going undercover'', as in certain forms of police and journalistic investigation that are important for the well-being of society. Meanwhile, authors have throughout history adopted the names of established authorities to enhance the influence of their own works. For example, Pseudo-Dionysius portrayed himself as Dionysius the Areopagite, the Athenian convert of Saint Paul. In modern times, celebrities frequently do the reverse, viz., employ ``ghostwriters'' to write books under their real names.

The journalist Lydia Laurenson defends pseudonymity in particular as a positive social practice. In her view, pseudonyms enable an ``elastic'' sense of self that ``is crucial for people’s personal development [...] We need space to experiment and risk-tolerant environments where people can learn'', environments she argues the Internet can, should and, until recent years, did provide~\cite{elastic_self}. Pseudonyms also provide camouflage for religious, ethnic, and sexual minorities, for whom experimentation, self-discovery, and self-expression, depending on their personal or social-political contexts, could be matters of life or death. Indeed, as platforms have been moving toward establishing a ``real-name Web''~\cite{real_name_web1}, there has been push-back from civil society in favor of pseudonyms --- effectively, in favor of fake accounts~\cite{real_name_web2}

By definition, inauthenticity seems irrelevant to pseudonymity, while relevant to allonymity. A mask does not necessarily purport to accurately represent anything about the underlying reality of its wearer, simply to hide their true identity, whereas imposture \emph{does} purport to accurately represent an underlying reality. Yet, complications arise even here. Consider the following thought experiment of a ghostwritten book about a celebrity's ``life philosophy''. On the one hand, the ghostwriter has done all of the actual writing and editing, while the celebrity has contributed neither. On the other hand, prior to writing the book, the ghostwriter extensively interviewed the celebrity, and even uses transcript elements in the finished text. This hypothetical book poses a conundrum: there seems some inevitable and crucial degree of inauthenticity, but on the part of whom, the ghostwriter or the celebrity?

Thinking closely about inauthenticity and pseudonymity reveals the complexities of deception. While \emph{all} fake accounts deceive, in principle not all fake accounts deceive \emph{maliciously}. It is malicious deception --- deception with the intention to commit some kind of harm against a target to the benefit of an adversary --- that is therefore key. Consequently, while platforms and everyday users have good reason to be fearful of fake accounts, they must also take care that they do not misinterpret their deception as invariably harmful. In other words, what is important is the \emph{intention} behind the act of creating a fake account. Fundamentally, the intention can be either malicious or non-malicious, even benign. 
\section{Thinking Taxonomically about Fake accounts}
\label{sec:taxonomy}

In this section, we offer two proposals: i) a definition of fake accounts and ii) four general aspects of fake accounts, each corresponding to one of the false dichotomies from above. 

Taxonomical thinking is a meta-level activity that needs to occur prior to, or at least simultaneously with, the crafting of a taxonomy. Before or during thinking about the taxa themselves, the would-be taxonomist must consciously think through the \emph{meta}-taxa, viz., the quantifiable \emph{and} the qualitative aspects --- the empirical elements and the intuitions --- that the eventual taxa are intended to capture. To borrow the notion of features-engineering from machine learning, the taxonomist must not only taxon-engineer, but also aspect-engineer.

The taxonomist must also definition-engineer. That is because every taxonomy entails an ontology, i.e., an answer to the ontological question, \emph{What is X?} In the case of fake accounts, any taxonomizer must first think through \emph{two} ontological questions -- what is fakeness and what are fake accounts? 

\paragraph{Definitions}

From the explorations of the foregoing sections, it is evident that fakeness is a form of deception in which there is a purposeful disconnect made by an agent between the outward reality of someone and the inward reality which it purports to represent. We define fake accounts as digital entities in which the act of fakeness --- again, understood as a form of deception --- is central to their nature. Hence, an account can be said to be ``fake'' if \emph{either} i) the entity with the alluded identity did not really create and/or control the account (allonymity), or ii) the alluded identity is fabricated and hence cannot be attached to any real entity (pseudonymity). Our definition of fake accounts extends beyond people to include what philosophers call ``social facts'' --- abstract but no less real mental objects, including concepts, games, words, gods, currencies, streets, neighborhoods, governments, religions, etc.~\cite{Searle}. In this way, we attempt to leave open conceptual space for future discussions about satirical accounts and other permutations of the fake account phenomenon. 

To be clear about an important nuance, by invoking the concept of ``entity'', we mean persons, objects, or concepts established by either formal and informal means. For example, an account falsely purporting to be a gamer from a particular gaming community is just as fake as an account using the identity of a citizen with a registered national identification number, passport, etc. Similarly, an account falsely purporting to be a real municipality is just as fake as an account falsely purporting to be a location from a science fiction novel.

Both definitions here do not immediately impugn malicious intent onto the agent. Whether a given fake account has a harmful effect on an audience is important to consider when attempting to taxonomize it. However, the existence of a harmful intention or effect should not be \emph{presumed}.

\subsection{Aspects}

With these definitions in hand, which aspects of fake accounts should be considered in the future when researchers, platforms, policy-makers, and others attempt to craft taxonomies? We propose the following four-part comprehensive framework that incorporates quantitative and qualitative modes of analyses addresses the four false dichotomies from above:

\begin{enumerate}
\item Accounts' scale, including quantity, distribution, whether coordination exists between them, etc. (corresponding to the false dichotomy of coordination versus non-coordination).
\item Accounts user(s), viz., person, program, or both (corresponding to the false dichotomy of programs versus persons).
\item Accounts' purpose(s) and technique(s) (corresponding to the false dichotomy of deception versus forthrightness).
\item Accounts' impact(s) on and implication(s) for their audiences (corresponding to the false dichotomy of inauthenticity versus authenticity).
\end{enumerate}

Note that our aspects themselves entail internal taxonomies, e.g., ``armies'' and ``legions'', program, person and cyborg,  tactics and behaviors, and so on. This means there is a kind of second-order meta-level of taxonomical thinking \emph{on top of} the first-order meta-level that we are elaborating here. Nevertheless, the would-be taxonomist need not descend into an \emph{infinitum ad absurdum}; rather, being able to think laterally is all that is required.

\subsection*{Aspect 1: Scale}

This aspect fundamentally concerns whether an ``army'' of fake accounts can be said to exist, comprised either of bots, sockpuppets, and/or other types of fake accounts~\cite{chinese_sockpuppet_army}. By ``scale'' we mean not only a) the sheer quantity of fake accounts, but also b) their distribution across a digital space \emph{as a network}, as well as c) how they coordinate. We further propose that an ``army'' of fake accounts $A$ can be said to exist if $ |A| > \tau$ where the determination of $\tau$ is variable because the figure should be determined by context. If the value of $\tau$ is high, this can enable a conversation about intermediate groupings of bots (such as whether ``legion'' as opposed to armies exist).

In general, the aspect of scale encompasses many of the directly observable ``structural'' behaviors of an account, e.g. the amount of connections an account has. Consequently, the aspect of scale can also be said to encompass various platform user and audience statistics.

However, scale cannot be properly reasoned about by only considering such statistics and we must consider whether and to what extent accounts act as singular versus multiple and simultaneous versus sequential broadcasters of content in order truly understand the scale. To illustrate this point, take the extreme example of a large set of Twitter bots who only connect with each other vs. a smaller set of Twitter bots that are distributed throughout the network. 

Similarly, collaboration is critical, as a well-coordinated smaller botnet may be more effective than a larger loosely collaborating one. Coordination can be via automation and/or direct human management~\cite{CyberSafety2020}, or a looser collaboration entailing complex networks of actors~\cite{starbird}. 

\subsection*{Aspect 2: User(s)}

This aspect concerns whether the immediate originator and/or controller of the fake account is a program or a person. Of course, programs ultimately serve people. For this reason, we emphasize the entity that is \emph{directly} either a) the source of the account's existence (``originator'') and/or b) is controlling it on a regular basis during a given time period (``controller''). Among other things, thinking in terms of originator versus controller helps to finesse the Ship of Theseus-style quandary we have described above with respect to the false dichotomy of program versus person.

This question of user intersects with the aspect of scale. An automated originator is the most likely source for bot armies, as a program can manufacture accounts at a rate and quantity far faster than even large teams of people. However, whether the originator is automated has less impact upon issues of administration. As noted above in our discussion about sockpuppets versus bots, the administrative scalability of certain kinds of fake accounts depends upon the nature of the user: simply put, a person, or even a team of people, will have difficulties administering fake accounts beyond a certain amount.

Thinking about the originator and controller is not only a question of \emph{whether} the user is a program or a person, but also the \emph{extent} to which they are either. Thus, the aspect of user(s) enables a conversation about ``cyborgs'', i.e., mixed program-person accounts. This thinking leads to interesting and difficult questions: ``How often in a given time period is can (or should) an account be controlled by a program or a person?'', ``What percentage cyborg does an account need to be in order to be `fake'?'',  ``Does being a cyborg inherently mean an account is fake?'', 

The aspect of the user is important beyond taxonomical questions and enters more deeply into the arena of ethics and policy-making. Increasingly, there is a need among those involved in the struggle with fake accounts to discuss and determine whether it is \emph{inherently} problematic for an account to be automated, in whole or in part, all or some of the time.

\subsection*{Aspect 3: Purpose(s) and Technique(s)}

This aspect concerns two levels of thinking: $a$) the over-arching intention/goal of the fake account (``purpose''), $b$) and the manner by which it accomplishes that purpose (``technique''). These levels can be further parsed into $a_1$) strategy versus $a_2$) objective and $b_1$) tactic versus $b_2$) behavior. In general, one can conceive of two broad purposes --- deception and non-deception --- thereby corresponding to the notions of fake and real. However, since there are degrees or intersections of deception and non-deception care and reflection are needed in order to avoid a false dichotomy. 

The strategy $a_1$ is how the account aims to accomplish its purpose (deception or not), while the objective $a_2$ boils down to the agenda of the purpose (e.g. to propagate more effectively) an advertisement). The question of strategy is quite fundamental, as in principle determining the strategy of an account should also determine, or at least inform, whether it is fake. 

Meanwhile, the tactic $b_1$ concerns the means by which the account actually completes its objective and the behavior $b_2$ is what the fake account concretely \emph{does} in pursuit of its objective. If translated into taxa, the distinction here would probably be best articulated as macro-, meso- and microscopic typologies composed of accounts, patterns and actions, e.g., fake followers versus social bots (macroscropic --- tactics); sharing or posting the same content \emph{verbatim} or expressing the same sentiments, not to mention the content itself (mesoscopic --- tactics); and liking, sharing, rhetoric, hashtagging, etc. (microscopic --- behaviors). 

To complicate this aspect, accounts can change their purposes multiple times over the course of their ``lifetimes'', as in the Ship of Theseus example. There also seems to be no limit to the amount of possible strategies and objectives that an account can pursue, evidenced by the way in which bots can switch from one activity to another~\cite{theseusbots2}.

Nevertheless, the utility of at least \emph{thinking} about purpose(s) and technique(s) lies in mentally separating out purpose from concrete action. In doing so, a more precise scientific conversation may be fostered about the purpose and actions of detection and control. Thinking about purpose(s) and technique(s) can also help inform discussions about scale, user(s) and audience impact(s) and implication(s). For example, one would expect that if an agent's purpose is to deceive, their strategy for doing so is to manipulate public opinion for or against a policy, their tactic is to use a legion of sockpuppets, etc., all of which factors into whether the originator of the accounts should be a program --- so as to reach the scale of a legion --- while the controller a person --- so at to more capably persuade the target audience, etc.

Although the purpose of this section is to propose aspects and not methods of detection and analysis, nevertheless the purpose of an account is difficult, if not impossible, to infer by quantitative methods. Therefore, such evidence may be gleaned from either the account's impact(s) on and implication(s) for its affected audiences, or by using features derived from its techniques as indicators.

\subsection*{Aspect 4: Audience Impact(s) and Implication(s)} 

This aspect concerns an account's impact on and implications for its target audience. By ``audience'' we mean two overlapping ideas: a) the traditional core/periphery model of social network analysis~\cite{core_periphery}, but also more loosely b) all those accounts that are directly connected to a given account via both connections and regular interactants with its content. Some consideration must also be given to those accounts that are second- or third-degree connections to an account  (friends of friends), as they are connected to it as irregular interactants with its content. Meanwhile, by ``impact'' we mean the cognitive, affective, and organizational dynamics of the target audience, while by ``implications'' we mean the short-, medium- and long-term ripple effects of these dynamics for either the target audience or other audiences connected to it.

Exactly \emph{which} impacts and implications should be taken into consideration is itself a crucial taxonomical question. For the sake of initiating the necessary conversation about this question, we propose as an (admittedly quite obvious) set of impacts ``discord'' versus ``solidarity'', and similarly as a set of implications whether and to what extent a target audience turns inward into so-called ``echo chambers'' or ``filter bubbles'' \emph{as a specific consequence} of fake account activity. Moreover, a further implication would be whether and to what extent, when an audience does turn inward, how does this withdrawal subsequently affect those audiences in closest proximity.

There is also an important question of whether the definition of audience here should be restricted to accounts, or if the project of grappling with fake accounts would be better served to include people in the ``real world'' --- and if so, to what extent? In a sense, this is already presumed by the very fact that fake accounts are experienced as a societal threat: if all that was stake was simply the integrity of digital spaces and there were no spillover effects into analogue spaces, perhaps the perception that fake accounts have reached historic crisis levels would not exist. Nevertheless, we argue that it is worthwhile to explore this question \emph{consciously} and from a taxonomical perspective. 

The aspect of audience impact(s) and implication(s) intersects and interacts deeply with the other proposed aspects above. The scale, user(s), and purpose(s) and technique(s) of fake accounts under question will obviously be a significant determining factors in both the quality and quantity of an account's impact and the ripple effects. The reverse also holds: one can make important inferences about the scale, user(s) and purpose(s), and technique(s) of a set of fake accounts by examining the dynamics of the audience before, during and after their activities.

All that being said, serious intellectual care must be taken with the aspect of audience impact(s) and implication(s). As was evident in the thought experiment above about the Wiggins' sockpuppets, there can be gaps in causation between the other aspects and this one. Recall that the Wiggins' audiences either did eventually learn of the deceit (Card's original scenario) or only suspected it (our revised scenario), but in both cases, they decided that the content of the Wiggins' fake conversation trumped the fact of the deceit.

\section{Taxonomical Thinking in Practice}
\label{sec:inpractice}
In the previous section we introduced a way of thinking taxonomically about fake accounts. Doubtlessly, our framework will prove to be incomplete in practice; researchers and others will need to revise or add to our proposed aspects, or develop entirely new frameworks of their own. That being said, in this section we explore how these aspects are expressed in practice by examining prior work on fake accounts, primarily focused on machine learning-based detection methods. 

Recently, research into the phenomenon of fake accounts has rapidly expanded, primarily due to the influx of fake accounts and the ease of Twitter data collection. Because of the speed of this increase, much of the research in this arena has been done --- out of necessity given the widespread impact of this problem --- without much taxonomical thinking. Our fundamental argument in this paper is that is necessary to think taxonomically from the very beginning of a study, as doing so (or \emph{not} doing so) affects the results at each level of analysis, starting even from the choice and implementation of data collection methods. For this reason, we will not only be critiquing, but also showing examples where taxonomical thinking helps to illuminate the work, both strong and weak, that researchers have done.

\subsection{Scale in Practice}

In the previous section, we discussed scale not solely as the quantity of accounts, but also in relation to their distribution and their methods of coordination.

First, in any classification task an estimate of the base rate (i.e. quantity) is necessary for precise classification. Without it, a na\"ively-trained classifier will perform well in validation, but fail in real world scenarios(e.g.~\cite{rauchfleisch2020false}). This is exceptionally difficult to estimate in the case of fake news classification, where the prevalence of fake accounts is unknown even to the platforms. Prior work has handled this issue in different ways. Some set the base rate arbitrarily ~\cite{el2016fake,benevenuto2010detecting} while others assume an even divide and handle the false positives post-hoc~\cite{yang2014uncovering}. In their 2015 paper, Cresci et al.~\cite{cresci2015fame} acknowledge this problem and solve it by testing across different base rates by under-sampling the positive class. While this method does allow for the base rate to be any value and thus introduces some flexibility, the base rate, and hence scale, is still needed in practice.

Related to quantity is distribution: not only how many there are, but where they are situated in the network graph. Without some knowledge about the distribution of fake accounts in the network, the data collection and ground truth labelling will be biased. Take for example a data crawler looking for spam accounts that begins with a seed set and finds other spam accounts that follow this seed set, then adds these accounts to the seed set and repeats. This operates on the (perhaps faulty) assumption that spam accounts that work in collaboration are friends or followers of one another. As such, the dataset will find spam accounts that work together, but it will not find the subset of accounts that do not rely on having contact with other spam accounts. Moreover, the classifier will be trained to find accounts with this pattern only, and thus it may miss non-friend/follower spam accounts if deployed outside of a lab setting.
 
Finally, we know from prior research, both academic and journalistic, that collaboration is common among fake account schemes and can be either automated~\cite{tj} or manual~\cite{CyberSafety2020}. This coordination should be considered not only in terms of how the accounts interact within the OSN itself, but also the coordinating activities that are occurring \emph{off} of the OSN. In other words, how fake accounts are coordinated both in the digital and analogue worlds should be considered~\cite{CyberSafety2020}. Without considering how the accounts coordinate offline, only those who interact publicly (e.g. via retweets) will be uncovered.

\subsection{User(s) in Practice}

We have witnessed not only a rise in the numbers of fake accounts on social media, but also their evolution as they develop new tactics to become more effective in their purpose and adapt to new detection methods. As such, there has been a diversification of the types of fake account schemes, including partially or fully manually controlled accounts that exploit cheaply available labor over automated computation. 

Unfortunately, some fake account detection methods miss these types of ``hybrid'' or ``troll'' accounts by falling into the false dichotomy that there are two types of accounts: bots and not bots~\cite{botornot,botsentinel}. Fortunately, researchers and platforms alike are beginning to understand the more complex configurations of accounts.

 \subsection{Purpose(s) and technique(s) in Practice}
 
Since most machine learning-based methodologies for fake accounts employ supervised methods of detection, a robust ground truth is needed. The methods for labeling ground truth vary, but most seek out technique as a proxy for purpose. That is, the methods of ground truth labeling begin not with the target accounts themselves, but with the activities that are assumed to be related to the purpose of the target accounts, e.g. fake accounts involved in spamming can be labeled as such by searching not for the accounts, but for spam itself~\cite{spam}. Similarly, fake accounts involved in astroturfing may be found not by searching for astroturfing accounts, but by detecting astroturfing attacks and labeling the accounts involved as fake~\cite{tj}. 

A common method for ground truth labeling is to assume suspended accounts were bots. This method rests on the assumption that accounts that have been suspended by Twitter are bots. This methodology has been effectively used for \emph{retrospective} studies to learn about the suspended accounts themselves, e.g. how many of the suspended accounts were able to gain normal followers~\cite{thomas11}, or how the suspended accounts participated in link-farming~\cite{linkfarming}.

This methodology has also been used to develop datasets to train machine learning models to detect new bot accounts~\cite{banlanangt,adaptivespammer}. While this may seem intuitive, it could also prove problematic. That is because the assumption that the suspended accounts were suspended because they were bots is only that --- an assumption. Perhaps it may seem like a reasonable assumption given the fact that the platform determined these accounts to have been bots. However, this begs the question of how the platform made that determination to begin with: did it take into serious consideration the purpose of the accounts, or did it suspend them based on a pre-formulated set of features?

\subsection{Audience Impact(s) and Implication(s) in Practice}

We have developed the aspect of audience impact(s) and implication(s) in part as a response to the false dichotomy surrounding inauthenticity. An essential facet of our critique of the latter is its moral presumption that every account is either completely fake or completely genuine, and with this, either malicious or benign. This is an assumption that does not hold up in the modern fake account ecosystem. 

Take for example a user who rents out their account to spammers or a political campaign in exchange for much needed cash~\cite{rent}, or a user whose account has been compromised in a password leak and unknowingly becomes part of a botnet~\cite{tj}. These cases complicate the idea not only of who the \emph{user} of an account is (and hence is relevant to that taxonomical aspect), but also complicates the more fundamental question of whether an account is fake, since at times it is genuine. As it is also difficult to ascribe malicious intentions onto users merely seeking to make a living, these cases further complicate the moral question of whether an account is malicious, since at times it is not, and even when it is, it was not intended to be. 

Perhaps more complicated still is the case of a user who sympathizes with a particular astroturfing campaign and retweets posts of bots. This is, often enough, the goal of an astroturfing campaign --- to gain traction within an audience. This user inadvertently, or even willingly, becomes part of the astroturfing campaign. Their intent may be malicious or benign, but this cannot be known. However, their actions are the same as a fake account controlled by an astroturfing algorithm.

\section{Conclusion}
\label{sec:conclusion}

This paper aimed to address the problem of a lack of conscious taxonomies for fake accounts. We presented a critique of how using classification as an implicit substitute for taxonomization leads to narrow and inaccurate pictures of the fake account ecosystem. We then identified and explored four false dichotomies that stem from binary thinking and that often come alongside primarily quantitative analyses of the phenomenon. We then addressed each of these false dichotomies by identifying and outlining four aspects of fake accounts by which one can think taxonomically about the phenomenon. 

Overall, we challenge the notion of inauthenticity and argue that intent is necessary in assessing fake accounts, however we also acknowledge that quantitative methods are unlikely to be able to determine intent, as there will never exist a clear signal or observable for intent to be correctly classified. That is, two users with different intents who take the same action are indistinguishable. As such, the only correct classifier would have as an input intention, however, this is impossible. Therefore, making decisions about interpretation purely based on classifier outputs is incomplete and hence faulty.

Because efforts to purge platforms of fake accounts are resource-intensive, determining the purpose and intentions behind them may be useful for gaining efficiency. Simply put, because not all fake accounts intend to cause harm, perhaps then not all fake accounts need to be eliminated. Beyond resources, we should also recon about whether accounts without malicious intention or negative effects on the network \emph{should be} taken down, and moreover, what is the role of the OSN in making these determinations. 

To conclude, recall from our discussion above the example of user who rents out her account to spammers or a political campaign in exchange for much needed cash. Cases like this may do more than complicate whether an account can be considered inauthentic, malicious, or fake, but complicate even more fundamentally whether it will continue to be intellectually valuable to distinguish between fake and real accounts at all. Given the fact that ascertaining the intention of a user is impossible (or, if it ever becomes possible, possibly unwise), it may instead increasingly prove prudent to bracket whether an account is fake and consider its current use at a single point in time: a single activity. 

If the passage of time reveals such bracketing to indeed be prudent, it may also eventually behoove researchers to examine more intensively the impacts on and implications for a given account's audience. We are indeed suggesting here that ultimately the very category of ``fake account'' itself may need to be dropped. While this may seem to undermine the very \emph{raison d'être} of this paper, the system of taxonomical thinking we have proposed above need not be restricted to fake accounts alone; instead, it can be expanded or revised to encompass accounts \emph{as such}. For this reason, the aspect of audience impact(s) and implication(s) may prove the most crucial for future research efforts into fake accounts, as ultimately it is this aspect more than the others which attempts to grapple with the concerns for audience integrity that underlie the concerns about fake accounts. Clearly, audience integrity is threatened not only by fake accounts, but by \emph{any} account that aims to cause harm. 

\section*{Acknowledgements}
This work is supported in part by the Open Technology Fund’s Information Controls Fellowship. The Civil Initiative for Internet Policies provides administrative, legal and logistical support. Special thanks to the team of journalists who helped out, comprised of Rustam Khalimov, Arina Efremova, Rauf Akhmatov and Aziza Isakova.

\bibliographystyle{plain}
\bibliography{citations}

\end{document}